\documentclass[aps,prb,twocolumn, showpacs,floatfix]{revtex4}
\usepackage{graphicx,epsfig,amsmath}
\usepackage{color}
\usepackage{soul}

\newcommand{\subfig}[2]{Fig.~\ref{fig:#1}(#2)} 

\newcommand{\Vtwo}{\mbox{$\text{V}_{\text{2}}$}}

\newcommand{\Tts}{\mbox{$\text{T}_{\text{2}}^{\text{*}}$}}

\newcommand{\Tplus}{\mbox{$\text{T}_{\text{+}}$}}
\newcommand{\Tone}{\mbox{$\text{T}_{\text{1}}$}}

\newcommand{\taum}{$\tau_m$}

\newcommand{\Amax}{\mbox{$\text{A}_{\text{max}}$}}
\newcommand{\Amin}{\mbox{$\text{A}_{\text{min}}$}}

\begin{document}

\title{The  visibility study of S-\Tplus \  Landau-Zener-St\"uckelberg oscillations without applied initialization \\ }

\author{G.~Granger}
	\affiliation{National Research Council Canada, Ottawa, ON Canada K1A 0R6}
\author{G.~C.~Aers}
	\affiliation{National Research Council Canada, Ottawa, ON Canada K1A 0R6}
\author{S.~A.~Studenikin}
	\affiliation{National Research Council Canada, Ottawa, ON Canada K1A 0R6}
\author{A.~Kam}
	\affiliation{National Research Council Canada, Ottawa, ON Canada K1A 0R6}
\author{P.~Zawadzki}
	\affiliation{National Research Council Canada, Ottawa, ON Canada K1A 0R6}
\author{Z.~R.~Wasilewski}
	\affiliation{National Research Council Canada, Ottawa, ON Canada K1A 0R6}
\author{A.~S.~Sachrajda}
  \email{Andrew.Sachrajda@nrc.ca}
	\affiliation{National Research Council Canada, Ottawa, ON Canada K1A 0R6}

\begin{abstract}

Probabilities deduced from quantum information studies are usually based on averaging many identical experiments separated by an initialization step. Such
initialization steps become experimentally more challenging to implement as the complexity of quantum circuits increases. To better understand the consequences of imperfect
initialization on the deduced probabilities, we study the effect of not initializing the system between measurements. For this we utilize Landau-Zener-St\"uckelberg
oscillations in a double quantum dot circuit. Experimental results are successfully compared to theoretical simulations.

\end{abstract}


\maketitle


\section{Introduction}

Spin qubits have generated a lot of interest recently in systems of single, double quantum dots (DQDs),\cite{Hanson2007, PioroLadriere2008,
Studenikin2012, StudenikinAPL2012} triple quantum dots (TQDs),\cite{Laird2010, Gaudreau2011, Aers2012, Studenikin2013, Medford2013} and two coupled
DQDs.\cite{vanWeperen2011}

Frequently probabilities are obtained from averaging thousands of individual measurements, with each measurement separated by an initialization pulse. For
example, to initialize into a singlet state a pulse can be applied to an appropriate location in the stability diagram where a dot electron will be replaced by one from the
leads with the appropriate spin (see, for instance, the supplementary information from Ref.~\cite{Petta2010}).
Such initialization pulses can also be used in larger TQD systems,\cite{Medford2013} but it is expected to be more difficult to continue implementing this technique in more
complex quantum dot circuits due to the isolation of the inner dot electrons. It is therefore important to understand the consequences and signatures of an
imperfect initialization step on the oscillation visibility. Experiments and numerical calculations aimed at the quantitative confirmation of these qualitative
predictions are therefore required. The observation that the visibility of coherent oscillations is qualitatively affected by the choice of pulse period in relation to the
spin relaxation time \Tone~is briefly discussed in Ref.~\cite{Nalbach2013}.

Here, we study the effect of not initializing the system between measurements. For this we utilize Landau-Zener-St\"uckelberg (LZS)
oscillations\cite{Landau1932, Zener1932, Stuckelberg1932, Shevchenko2010, Ribeiro2010, Burkard2010, Sarkka2011, Petta2010, Nalbach2013} in a double quantum dot circuit. In
Section~\ref{exp det}, we describe the sample and the experimental setup used in the measurements. In Section~\ref{LZS}, we explain the physics behind the LZS oscillations,
as this is crucial for the understanding of the main parts of the paper. Section~\ref{calc} shows the details pertaining to the model used for comparative theoretical
calculation. The experimental and theoretical LZS oscillations are compared in Section~\ref{comp}. The observation that the visibility of the oscillations can be optimized
is included in Section~\ref{visibility}, while a discussion of the effects due to the nuclear spins in the lattice is detailed in Section~\ref{DNP}.


\begin{figure}[tbh]
\setlength{\unitlength}{1cm}
\begin{center}
\begin{picture}(8,7)(0,0)
\includegraphics[width=8cm, keepaspectratio=true]{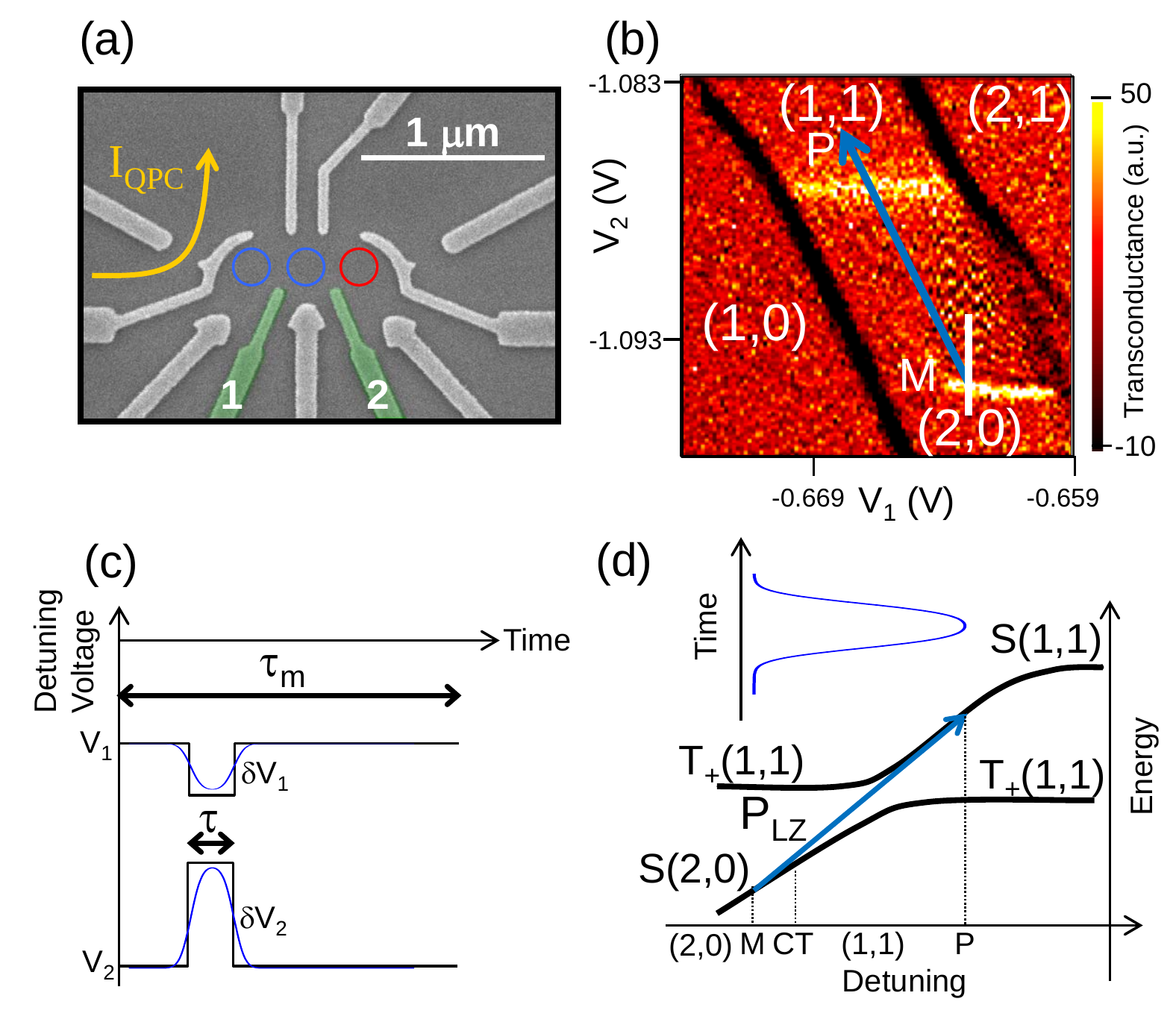}
\end{picture}
\end{center}
\caption{(Color online) (a) Electron micrograph of a device similar to the one measured. The gates define the triple quantum dot potential and two QPCs used as charge
detectors. The blue circles schematically indicate where the dots under study are located and the red shows the unused dot. Gates 1 and 2 receive short voltage pulses
shown in (c) in addition to DC voltages $V_1$ and $V_2$. The left QPC current $I_{QPC}$ is used to detect the electron number,  plot the stability diagram, and perform the
spin readout. (b) Stability diagram from the measured transconductance of the left QPC in the V$_1$-V$_2$ plane. The transconductance is the numerical derivative of the
left QPC conductance with respect to $V_2$. The electronic configurations are indicated, and LZS fringes are seen in the (2,0) region. The pulse size and direction are
indicated by the blue arrow. M is the measurement point, while P is the maximum detuning point during the pulse. ($\delta V_1,\delta V_2$)=(-5,10)~mV, $\tau$=16~ns,
\taum=2~$\mu$s, and B=80~mT. The rise time is 8~ns. The initial detuning line relevant to the data in \subfig{3}{a} is shown with a white line.
(c) Schematic showing the definition of the rectangular pulses (black) and the resulting Gaussian-convoluted pulses used here (blue). $\tau$ is the pulse duration, while
\taum~is the pulse period. The pulse amplitudes $\delta V_1$ and $\delta V_2$ applied to gates 1 and 2 are also indicated prior to the convolution. It is expected that for
the rise times used here, the effective amplitude will be a little smaller than what is shown on the rectangular pulse. (d) Schematic of S-\Tplus~energy diagram as a
function of detuning showing the anticrossing between S and \Tplus. During the pulse, there is a Landau-Zener probability P$_{LZ}$ of reaching the upper branch of the
S-\Tplus~anticrossing. ``CT'' indicates the charge transfer line. Inset: Gaussian-convoluted pulse.}
\label{fig:1}
\end{figure}

\section{Experimental details}
\label{exp det}

The device geometry is shown in \subfig{1}{a}.\cite{Gaudreau2009, Gaudreau2011} It is fabricated on a GaAs/AlGaAs heterostructure with a two-dimensional electron gas
(2DEG) located 110~nm below the surface with a density of 2.1$\times10^{11}$~cm$^{-2}$ and a mobility of 1.72$\times10^6$~cm$^2$/Vs. TiAu gate electrodes, patterned by
electron-beam lithography, define the quantum dot potential profile and quantum point contacts (QPCs).\cite{Field1993} Charge detection measurements are made with the left
QPC, tuned to a conductance
regime very sensitive to the local electrostatic environment, below 0.1~e$^2$/h, using a lock-in technique with a typical 50~$\mu$V rms modulation. Short DC (rectangular)
pulses of duration $\tau$ (defined prior to the Gaussian convolution) and period \taum~from two synchronized arbitrary waveform generators (Tektronix AWG710B) are applied
to gates 1 and 2 via bias-tees to quickly change the dc voltages $V_1$ and $V_2$ by small increments $\delta V_1$ and $\delta V_2$.  As \taum~is orders of
magnitude larger than $\tau$, \taum~is called the measurement time in practice. The \taum~values will be varied in such a way to span the regime where the measurement time
is much longer than the spin relaxation time \Tone~(so the next initial state is the ground state) to the regime where the measurement time is smaller than \Tone~(so
incomplete relaxation provides either the excited state or the ground state as the next initial state). The pulse rise times are controlled by loading pulses that
correspond to the numerical convolution between a rectangular pulse and a Gaussian function into the arbitrary waveform generators [\subfig{1}{c}].   The experimentally
measured resulting rise times are defined in the region from 10\% to 90\% of maximum amplitude. The magnetic field is applied parallel to the 2DEG. To reduce
telegraphic noise issues the device is bias-cooled with 0.25~V on all gates from room to the base temperature of the dilution refrigerator.

We focus on the (2,0) and (1,1) regions of the stability
diagram [\subfig{1}{b}], where  (N$_L$,N$_C$) represents the electronic configuration with the indicated number of electrons on the left and center quantum dots
respectively. The device is a triple quantum dot lithographically; however, the measurements are made in a regime where the right dot is detuned away from this part of the
stability diagram and plays no role in the measurements. The device therefore is an effective double quantum dot and the center quantum dot will be referred to as the
right dot from now on.

The thick black lines in \subfig{1}{b} are addition lines; the electron number in one of the dots changes by one whenever one of these lines is crossed. The upper yellow
horizontal line in \subfig{1}{b} is the charge transfer line between the (2,0) and (1,1) regions; whenever this line is crossed, an electron is  transferred
from the left dot to the right dot (or
vice versa). The other features, including the lower yellow line, pertain to the presence of the pulse and will be described in the next section.

\section{Landau-Zener-St\"uckelberg oscillations}
\label{LZS}

The experiments involve the creation of a superposition by passage through an anticrossing (analogous to a beam splitter), phase evolution, completing the interferometer by passing through the anticrossing a second time and then reading out the state.
 The required anticrossing occurs naturally in the scheme. The state of two electrons in the (2,0) or (1,1) electronic configurations can either be a spin 0 singlet or a
spin 1 triplet. As a result of the magnetic field applied along the z direction and the sign of the electron g-factor in GaAs, the lowest energy triplet
component is \Tplus=$|\uparrow,\uparrow>$. There exists a point in the (1,1) region of
the stability diagram along the pulse detuning line where the S and \Tplus~states anticross. This is illustrated schematically in \subfig{1}{d}, where we
ignore the other triplet components and the S(1,1)-S(2,0) anticrossing to focus on the particular anticrossing in which we are interested. For the full state spectrum the reader is referred to Ref.~\cite{Petta2010}.  
This anticrossing between the S and \Tplus~states originates from
spinflips mediated by the hyperfine interaction between the electron spins and the host lattice nuclear spins.\cite{Petta2010}
 The confined electron spins experience an effective Overhauser magnetic field originating from the lattice nuclei. This is slightly different for electrons in the two dots, purely from statistical considerations, resulting in Overhauser field gradients. The interaction at the anticrossing is proportional to $g\mu_B\Delta B_{x,y}$, where $g$ is the g-factor of conduction electrons in GaAs, $\mu_B$ is the Bohr magneton, and $\Delta B_{x,y}$ is the Overhauser field gradient perpendicular to the applied magnetic field.

The initial diabatic transition probability between the two eigenstates of the S-\Tplus~anticrossing, when applying a pulse which detunes the
system through the anticrossing from (2,0) to (1,1), is given by
\begin{equation}
\mbox{P}_{LZ}=\mbox{exp}\left\{-\frac{(2\pi g \mu_B \Delta B_{x,y})^2}{h|\frac{d}{dt}(E_+-E_-)|}\right\}
\end{equation}
where $h$ is Planck's constant and $E_+$ and $E_-$ refer to the eigenenergies of the two states involved in the S-\Tplus~anticrossing.\cite{Petta2010} The key to creating
a quantum superposition between S and \Tplus~is to allow the detuning to change at the appropriate speed, for which the Landau-Zener probability differs from
zero. If the detuning changes extremely slowly, $P_{LZ}=0$ as for an adiabatic process, while if the detuning changes extremely fast, $P_{LZ}$=1 as for a fully diabatic
process.

Once a quantum superposition is created, the relative phase difference $\Delta \phi$ between the two eigenstates is given by the time integral of the
detuning-dependent energy difference between these states, i.e.~$\Delta \phi=\int_0^t{(E_+-E_-)\mbox{d}t'}/\hbar$. At fixed initial detuning, increasing the
pulse duration would lead to oscillations in the probability of finding either spin state, as $\Delta \phi$ increases linearly with the time spent at the end of the pulse
in the (1,1) region. In this paper we vary the initial detuning while holding the pulse duration constant. The pulse duration of 16~ns was determined mainly
by limitations due to the decoherence time \Tts~(which lies in the 5-15~ns range here) and the chosen pulse rise time (8~ns). The formula for $\Delta \phi$
still applies in the case of fixed pulse duration. As the initial detuning increases, the pulse goes further in the (1,1) region past the
S-\Tplus~anticrossing, the energy difference $E_+-E_-$ increases, and therefore the relative phase $\Delta \phi$ grows. This leads to oscillations in the probability of
finding either spin state as a function of detuning. This describes the phase accumulation near the top part of the pulse, at largest detuning. This process needs to be
integrated everywhere the pulse is past the anticrossing. Ideally $P_{LZ}$ would be $\sim$0.5 for optimum superposition. For the pulse and statistical
inhomogeneous field gradient in these experiments, the initial value for $P_{LZ}$ after passing through the anticrossing is, however, approximately $\sim$95\%. The final
part of the applied pulse detunes the system back through the anticrossing. $P_{LZ}$ will once again change the quantum superposition of the already superposed states.  In
theory, it would be possible to apply a more complex train of pulses prior to commencing the phase evolution experiment to improve the degree of initial superposition, but
the experiments and theory in this paper involve superpositions created by single pulses.

In the case of the S-\Tplus~anticrossing, the charge detector cannot distinguish the two spin states in the (1,1) region, as the underlying charge state is the same for both states. However, on passage through the anticrossing a second time and into the (2,0) region of the stability diagram, the S component evolves smoothly to the (2,0) ground state, while \Tplus~ remains in the excited state with a (1,1) charge configuration since the triplet (2,0) is not energetically accessible. Thus after the pulse is complete, at the measurement point M in the (2,0) region,  the resulting spin state information can be converted into charge information, a process referred to as spin-to-charge conversion,\cite{Hanson2007, Ono2002} as the spin states map to different charge states there. The charge detector (a nearby quantum point contact) makes a measurement by effectively  projecting the spin state on the S and \Tplus~basis states. In practice, the QPC conductance is averaged over $\sim$10$^5$ identical pulses. The probability of returning in S after such an experiment depends upon many factors such as the effectiveness of the Landau-Zener process, the relaxation time, and decoherence time of the system.

 Certain reported experiments have utilized an initialization step before every measurement in the averaged collection (e.g.~Ref.~\cite{Petta2010}) by adding a preliminary pulse toward one of the addition lines in the (2,0) region prior to the detuning pulse described above. This allows for an exchange of an electron between the dot and the lead to achieve the desired singlet state. We call these IS experiments to differentiate them from experiments where there is no such pulse (NIS). In the latter case the resulting state at the end of the measurement is used as a starting point of the next pulse. The statistics over many pulses will still oscillate as a function of detuning. In this paper, we compare these two procedures.

We can infer the position of the S-\Tplus~anticrossing in the stability diagram from the position of the lower horizontal yellow line in the (2,0) region of
the stability diagram in \subfig{1}{b}. When the measurement point M is on this yellow line, the tip of the pulse P reaches the S-\Tplus~anticrossing. Above this yellow
line are fringes corresponding to LZS oscillations since in that regime the detuning pulse passes through the S and \Tplus~states anticrossing in the (1,1) region [see
\subfig{1}{b,d}]. The resolution is poor due to a small number of averages, but  is sufficient to locate the region where LZS oscillations are present. Note that, in the
stability diagram of \subfig{1}{b}, the QPC conductance data  are numerically differentiated with respect to the \Vtwo~gate voltage for an improved presentation.  In all
other graphs the QPC conductance data are normalized relative to the conductance step at the charge transfer line between (1,1) and (2,0) to get a probability of return in
state S between 0 and 1.

\begin{figure}[hbt]
\setlength{\unitlength}{1cm}
\begin{center}
\begin{picture}(8,10.5)(0,0)
\includegraphics[width=8cm, keepaspectratio=true]{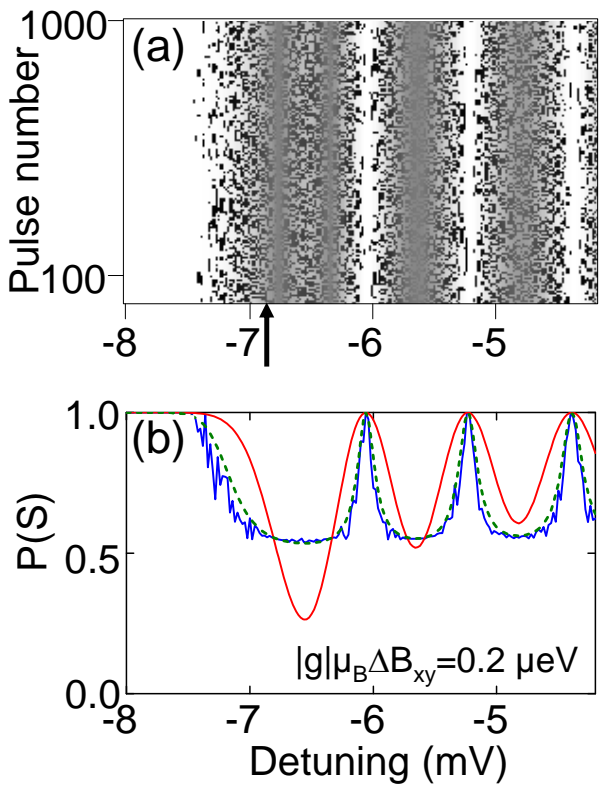}
\end{picture}
\end{center}
\caption{(Color online) (a) Calculated 2D map of P(S) vs.~initial detuning for 1000 consecutive pulses. Black is P(S)=0 and white is P(S)=1. The arrow indicates where the S-\Tplus~anticrossing is observed on the initial detuning axis (as in the lower horizontal yellow line in \subfig{1}{b}). (b) Calculated P(S) vs.~initial detuning for hyperfine splitting $|g|\mu_B\Delta B_{xy}$=0.2~$\mu$eV of the IS case
(red). The blue curve is the NIS case from the average over 1000 pulse repeats. The green dashed curve is NIS derived from IS (see text).}
\label{fig:2}
\end{figure}

\section{Calculations details}
\label{calc}

Throughout the paper we distinguish data with an initialization step and without initialization step. Let us explain what this means from the point of view of the calculations and the experiments. In the calculations, No Initialization Step (NIS) means that the final result for P(S) is the probability averaged over 1000 pulse sequences, where the final state (S or \Tplus) determined after any given pulse is the starting state for the next pulse. In the calculation, what we call the Initialization Step (or IS) is just the resulting P(S) after the first pulse, where the initial conditions prior to the pulse have been set theoretically to the singlet state. Experimentally, No Initialization Step (NIS) means we only apply the Gaussian-convoluted pulses and no other pulses.\cite{Footnote}

The NIS repeat pulse scheme is simulated theoretically
with the following model based on the solution of the time dependent density
matrix.\cite{Gaudreau2011, Taylor2007} The system is initialized in the S state
after which the first pulse is applied. During the measurement cycle the
population is allowed to partially relax to the ground state with a decay time \Tone. At the
end of each pulse cycle the state vector is then projected onto the charge basis to
yield either S or \Tplus~occupation, which then defines the initial state for the
next pulse cycle. As in experiment, P(S) is obtained as the fraction of measurements that find the 
system in the S state. Typically 1000 such cycles provide sufficient
statistics to model the effects of relaxation and initial state variation (typical experimental
sequences involve $10^{5}$ pulses).
Since the measurement time after each pulse had a duration much greater than the pulse itself
it was found impractical to include the measurement segment in the simulation. By the same token, however, 
almost all of the relaxation occurs during this measurement time when no other relevant time variation is present. Thus, the $\Tone$ 
decay was not included explicitly in the density matrix and was instead added by means of 
an exponential term prior to each projection operation.

A typical 1000 cycle repeat of the calculated P(S) vs.~initial detuning scan is shown in \subfig{2}{a} for
a low value of \taum/\Tone$\sim$0.1, where consequences of relaxation are less important, and a hyperfine splitting $|g|\mu_B\Delta B_{x,y}$=0.2~$\mu$eV,
consistent with experimentally measured values for the hyperfine splitting.\cite{Gaudreau2011}
For a given initial detuning, the vertical axis shows the result of each
consecutive measurement. For the same parameters, \subfig{2}{b} shows the IS case from the first pulse response in red, and the NIS case from the averaged probability over
1000 pulses in blue [averaging over the vertical direction in \subfig{2}{a}].

 It is instructive to derive the NIS result from the single pulse (IS) curve for
P(S) using an iterative approach. Here P(S) serves as the starting estimate of
the NIS probability of being measured in S. Due to symmetry it is also the
probability, we call it P in this role, of a single pulse leaving the system
in the state it started in, whether that state be S or \Tplus. 
As can be seen from Eq. (1), the probability of diabatic transition
depends only on the time derivative of the energy spacing and not on the choice of initial state (S or \Tplus).
Likewise (1-P) is
the probability of a single pulse switching the final state from that in which it
started. Under NIS conditions we can write a self-consistent condition for the
total probability of measurement in the S state as the sum of the probability of
starting in the S state and staying there after another pulse plus the
probability of starting in the \Tplus~state and then switching during the pulse:
\begin{equation}
\mbox{P}(\mbox{S}) = \mbox{P}'(\mbox{S})\mbox{P} + [1 - \mbox{P}'(\mbox{S})](1 - \mbox{P})
\end{equation}
where $\mbox{P}'(\mbox{S}) = 1 -  [1 - \mbox{P}(\mbox{S})]$$\mbox{e}^{-\tau_m / \mbox{T}_1}$ accounts for the decay from \Tplus~to S over the pulse sequence.

This has an analytic solution for the NIS averaged probability of returning in  S as
\begin{equation}
\mbox{P'(S)} = \frac{\mbox{Pe}^{-\tau_m / \mbox{T}_1}-1}{(2\mbox{P-1)e}^{-\tau_m / \mbox{T}_1}-1}
\end{equation}

\noindent where P is the single pulse curve (red curve in \subfig{2}{b}). The iterative formulation in the above analysis is a mathematical construction not to be confused with pulse repeats. As can be seen in \subfig{2}{a}, the statistical distribution of the pulse measurements in the numerical simulations do not vary from pulse to pulse.
The above procedure gives the green dashed line in \subfig{2}{b} which agrees well
with the 1000 pulse repeat result. It also provides insight into the form of the
NIS curve. Note, for example, that the expression for P'(S) will always produce a value between 0.5 and 1.0, even from a P(S)
minimum below 0.5, as in \subfig{2}{b}. For P=1 we get  P'(S)=1; for P=0 we get P'(S)=1/(1+$\mbox{e}^{-\tau_m / \mbox{T}_1}$) which gives 0.5 for small \taum / \Tone~and 1 for large \taum /\Tone, i.e.~P'(S) lies between 0.5 and 1. More generally, when \taum/\Tone~ is small, the result is always 0.5 for any value of P not exactly equal to 1. Thus, a cosine-like form for the single pulse P leads to a non-cosine-like NIS probability peaking sharply and having values $>$0.5.

 The oscillations in the IS case of \subfig{2}{b} appear to become damped as initial detuning increases. This is due to the fact that $P_{LZ}$
is optimal to create a superposition only near the S-\Tplus~anticrossing. At detunings further away from the anticrossing, $P_{LZ}$$\sim$95\% and the superposition
contains less and less of \Tplus~hence the higher measured values of P(S). To first order, the detuning period of the oscillations, $T_\epsilon$ (in mV), along the
detuning axis (i.e. along the $\epsilon$ axis) is inversely proportional to $|{d(E_+-E_-)}/d\epsilon|$ at the peak of the pulse (where most of the phase difference is
accumulated). The detuning dependence of $E_+-E_-$ is shown in \subfig{1}{d}. For instance, over detuning ranges where the eigenstates change approximately linearly with
$\epsilon$,  the derivative will be a constant,  so $T_\epsilon$  is constant. Therefore, equally spaced fringes along the detuning axis are expected as a function of
detuning  where $E_+-E_-$ varies approximately linearly with $\epsilon$ (i.e. some distance away from the anticrossing).

The most prominent feature in the NIS case shown in \subfig{2}{b} is that in narrow regions around where
P(S)=1 in the IS case (red curve), the pulse averaged probability in the NIS case (blue curve) remains
near unity.
 This is because the singlet that is measured after the first pulse becomes the initial and final state for all subsequent pulses so P(S)$\sim$1.
 As the detuning is moved away from these points, the projection
operation introduces more statistical cases where P(S)=0, and, in the regions
where P(S) is a minimum in the IS case, the minimum in the NIS case broadens and approaches 0.5 at this low \taum/\Tone$\sim$0.1. For the first
minimum in the IS case that drops below 0.5, the probability in the NIS case is noisier but again averages towards 0.5 (the statistical average of these pure
\Tplus~and S results after each pulse leads to P(S)$\sim$0.5). This is true independently of detuning, so there is little appreciable damping in the P(S) minima along the
detuning axis in the NIS case.

\section{LZS oscillation amplitude dependence on pulse period}
\label{comp}

\begin{figure}[htb]
\setlength{\unitlength}{1cm}
\begin{center}
\begin{picture}(8,14.7)(0,0)
\includegraphics[width=8cm, keepaspectratio=true]{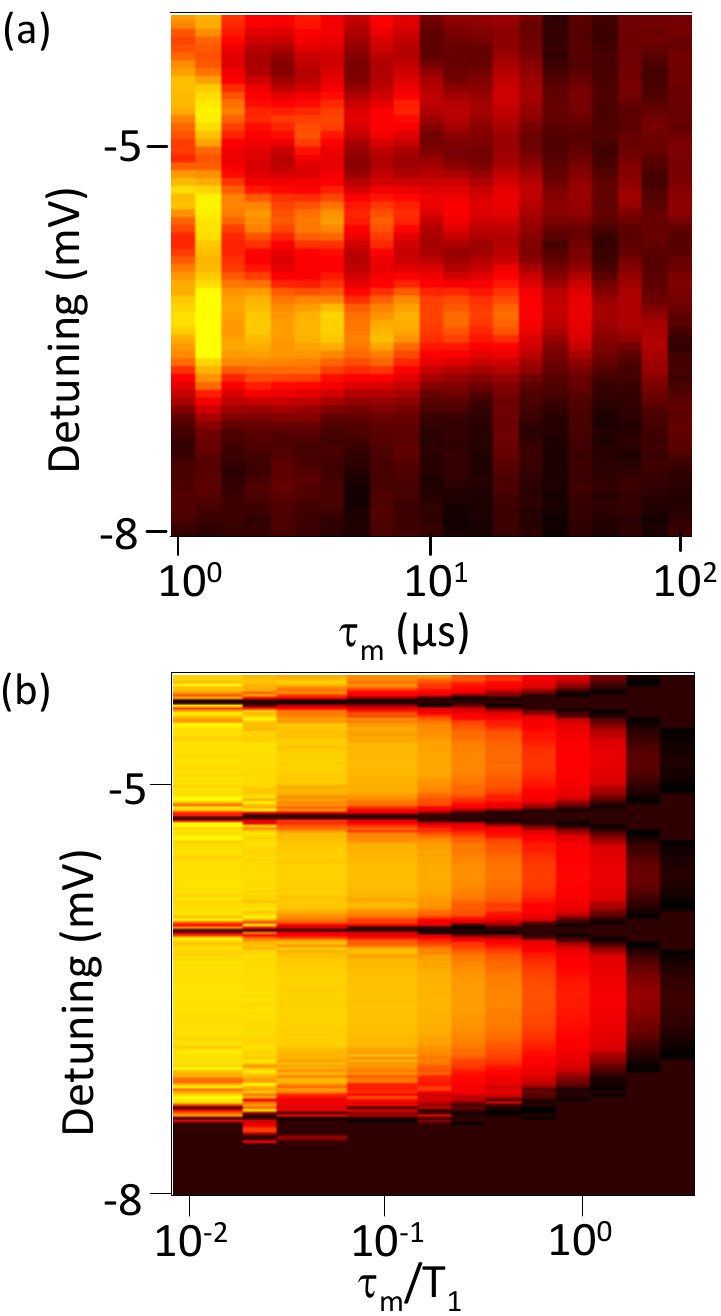}
\end{picture}
\end{center}
\caption{(Color online) (a) LZS oscillations of P(S) in the \taum-initial detuning plane from the  left QPC
conductance (plane subtracted) taken along the white line from \subfig{1}{b}. The initial detuning is with respect to the position of the charge transfer line in
\subfig{1}{b}.  ($\delta V_1,\delta V_2$)=(-5,10)~mV, $\tau$=16~ns, and B=80~mT. The rise time is 8~ns. Yellow is low, red is medium, and black is high. (b)
Calculated P(S) to compare to (a).}
\label{fig:3}
\end{figure}

The visibility of the LZS oscillations in P(S) is expected to depend on the choice of pulse period \taum~and on the value of \Tone~at the measurement
point.\cite{Nalbach2013}  A few periods of LZS oscillations vs.~initial detuning and \taum~are shown in \subfig{3}{a}. These oscillations are obtained by
using a single Gaussian-convoluted rectangular pulse of a constant amplitude.\cite{Petta2010, Gaudreau2011}  The peak amplitude of each LZS oscillation decreases as
\taum~increases due to the spin relaxation time \Tone.  We extract \Tone~as the value of \taum~where the peak amplitude decreases to 37\% of its maximum value. We find
that \Tone~ grows from 20~$\mu$s to 60~$\mu$s as initial detuning becomes more negative in \subfig{3}{a}. Because the spectrum of excited
states changes with detuning as the DQD potential is deformed by the gate voltages, the  inelastic decay mechanisms for the states also change, hence the triplet excited
state relaxation time \Tone~varies with detuning. 

Figure~\ref{fig:3}(b) plots the results of a calculation in the NIS case that corresponds to the experimental case of \subfig{3}{a}. Our calculation is based on
the model described above and in Refs.~\mbox{\cite{Taylor2007, Gaudreau2011}}.   In the calculation, the detuning dependence of \Tone~at the measurement point is not taken into
account. The value used is \Tone=60~$\mu$s.  This explains why the three calculated LZS peak amplitudes decay in an identical fashion versus \taum~in contrast to the experimental
results in \subfig{3}{a} where \Tone~does indeed vary with detuning.

The calculation in \subfig{3}{b} reveals that, at large \taum/\Tone, the
oscillations along the initial detuning axis are approximately sinusoidal (i.e.~peaks and dips have equal widths along the detuning axis) and that, at
small \taum/\Tone, the dips become much wider than the peaks, just as is observed in the experimental results of \subfig{3}{a}. Even though the corresponding features are smeared out somewhat in experiment (presumably because of decoherence effects not
included in the present model), the non-sinusoidal character of
the oscillations for smaller \taum~is clear, as one can clearly see that below \taum$\approx$10~$\mu$s the yellow regions are wider than the black regions in \subfig{3}{a}.

\begin{figure}[htb]
\setlength{\unitlength}{1cm}
\begin{center}
\begin{picture}(8,6.5)(0,0)
\includegraphics[width=8cm, keepaspectratio=true]{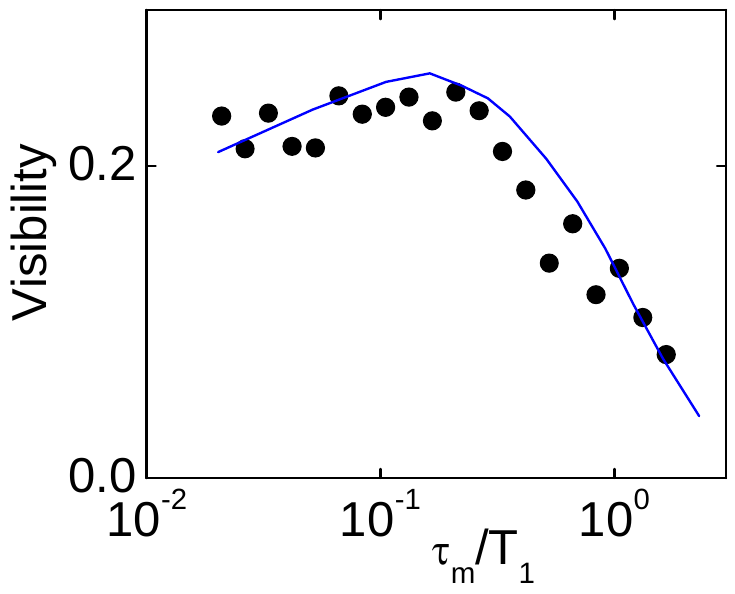}
\end{picture}
\end{center}
\caption{(Color online) Visibility as a function of \taum/\Tone~from the experiments in \subfig{3}{a}, where we assume \Tone=60~$\mu$s as a fitting parameter (filled circles) and NIS calculations (blue line). Calculation parameters are the same as for the calculation in \subfig{2}{a,b}; in particular, $|g|\mu_B\Delta B_{xy}$=0.2~$\mu$eV.}
\label{fig:4}
\end{figure}

\section{LZS oscillation visibility}
\label{visibility}
The oscillation visibility is a figure of merit with several possible definitions depending on the application. When there is little noise in the
oscillations, it is possible to take the definition of visibility as (\Amax-\Amin)/(\Amax+\Amin), where \Amax~and \Amin~are the extrema of the
oscillations. In the presence of noisy oscillations requiring an average of a large number of measurements, a statistical approach based on the
standard deviation is suitable to extract the visibility. In order to compare the calculation and the experiment, we use the same definition of
the visibility, $V$, in both cases: $V=2\sigma$, where $\sigma$ is the standard deviation of  P(S) calculated along a given detuning trace at a fixed value of \taum. An integral number of LZS
periods are included in the visibility calculation  (three in our case). We note that this definition of $V$ would coincide with the ratio (\Amax-\Amin)/(\Amax+\Amin) from the corresponding
ideal positive sinusoidal waveforms if the noise was subtracted.



Maps such as those of \subfig{3}{a} and (b) can be used to determine whether there exists an optimal value of \taum/\Tone~that maximizes the
visibility. The curve for the calculated NIS visibility vs.~\taum/\Tone, shown as a solid blue line in Fig.~\ref{fig:4}, can be understood as follows. At values
of \taum$>$\Tone, spin relaxation reduces the spin-blockade signal towards zero, hence the visibility of the LZS oscillations also decreases. At
\taum~much smaller than \Tone, not enough time is spent in the measurement phase at point M for spin relaxation toward state S to occur. The projection of the qubit state vector onto the
S-\Tplus~basis after the pulse therefore becomes the starting point for the next pulse. Even though the result may alternate statistically between P(S)=1
and P(S)=0, averaging over several pulses will lead to an overall P(S)$\sim$1/2 and to a reduced visibility by the same averaging process as in
\subfig{2}{b} (the standard deviation for the NIS case is smaller than that for the IS case). For the given experimental conditions (\textit{i.e.}
for specific interdot couplings), the calculated optimum visibility occurs at a ratio \taum/\Tone$\sim$0.2.


Experimentally, the visibility is also extracted as $2\sigma$ over three peaks of each detuning trace from \subfig{3}{a}. The results are plotted
as filled circles in  Fig.~\ref{fig:4} and match the prediction from the calculations. The visibility varies by $\sim$10\% depending on how many
periods are included in the visibility calculations because the amplitude varies with detuning as discussed above. The optimum predicted
theoretically is observed at \taum/\Tone$\sim$0.2, assuming \Tone=60~$\mu$s as a fitting parameter for the case presented in  Fig.~\ref{fig:4}.



\section{Discussion of effects due to dynamical nuclear polarization}
\label{DNP}

\begin{figure}[bth]
\setlength{\unitlength}{1cm}
\begin{center}
\begin{picture}(8,10.5)(0,0)
\includegraphics[width=8cm, keepaspectratio=true]{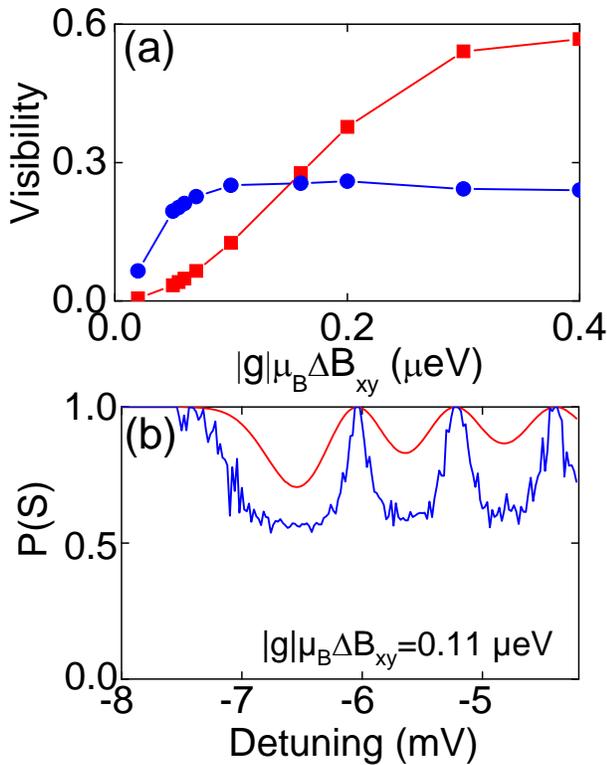}
\end{picture}
\end{center}
\caption{(Color online) (a) Calculated IS (red squares) and NIS (blue circles) visibility for \taum/\Tone$\sim$0.1 as a function of $|g|\mu_B\Delta B_{xy}$. (b) Calculated
P(S) vs.~initial detuning for hyperfine splitting $|g|\mu_B\Delta B_{xy}$=0.11~$\mu$eV of the IS case (red). The blue curve is the NIS case for comparison.}
\label{fig:5}
\end{figure}

Our calculations indicate surprisingly that it is possible for the visibility obtained for IS to be less than that obtained in
the statistical way for NIS.  Figure~\ref{fig:5}(a) shows the calculated visibilities $V$ obtained for the IS and NIS cases as a function of the $\lvert$g$\rvert \mu _B \Delta B_{x,y}$ hyperfine splitting responsible for the S-\Tplus~anticrossing. 
Since the value of P$_{LZ}$ strongly influences the magnitude of the P(S)
oscillations and the subsequent averaged NIS visibility and P$_{LZ}$ is directly
modified by the hyperfine splitting, we would expect to see significant
variations of the visibility with the splitting.
 For smaller splittings than usually occur in our experiments, the two curves cross and the NIS visibility is actually larger than that of IS. The calculated
P(S) for the two cases at $|g|\mu_B\Delta B_{xy}$=0.11~$\mu$eV are shown in \subfig{5}{b}. For this hyperfine splitting P$_{LZ}$ increases beyond 98\% at
detunings greater than
-7 mV leading to P(S) oscillations that are much reduced relative to \subfig{2}{b}.
In the NIS case, the averaged minima are pulled down toward 0.5 leading to enhanced visibility.
  In summary, the visibility of LZS oscillations does not necessarily increase due to standard initialization (it depends on the specific  experimental situation).

The visibility and the relaxation time can both be affected by dynamical nuclear polarization (DNP). The interplay between visibility, relaxation time, and pulse rise time has been investigated recently in a DQD, where DNP pulses were applied to increase the value of $\Delta B_z$ due to hyperfine nuclear interaction.\cite{Barthel2012} Even though we do not specifically apply a DNP pulse sequence here prior to the qubit manipulation pulse, we know from Ref.~\cite{Gaudreau2011} that this pulse results in a small DNP effect in our samples. DNP may also account for some visible difference between experimental and theoretical graphs in  \subfig{3}{a} and \subfig{3}{b}.
Even though DNP effects are beyond this study, the presented theory correctly captures the main experimental features discussed above.

\section{Conclusions}

In conclusion, we have carried out the experimental and theoretical study of the microscopic mechanisms affecting the visibility of  Landau-Zener-St\"uckelberg oscillations in conditions where an initialization step is difficult.  The results apply to cases where one of the dots is isolated from the leads (e.g. the center dot of a triple quantum dot) or where the relaxation time \Tone~is too short (as for charge qubits).
Partial initialization occurs in these situations due to the \Tone~relaxation process during the qubit readout step.

The visibility depends on the ratio of pulse period \taum~to the relaxation time \Tone; it reaches a broad maximum at an optimum point, which depends on system parameters.  In our experiment the LZS visibility reaches a maximum of 0.25 at the optimal  \taum/\Tone$\sim$0.2 in good agreement with theoretical calculations. Theoretical analysis shows that in some cases (for smaller hyperfine interactions) the visibility without initialization step can exceed the one with the initialization.
It is important to find optimal settings for \taum/\Tone~that can also be be used for the estimation of \Tone~or the level coupling at the anticrossing.

\acknowledgments

We thank O. Kodra for programming. A.S.S. acknowledges funding from NSERC Grant No. 170844-05. G.G. acknowledges funding from the NRC-CNRS collaboration.


\begin{thebibliography}{00}


\bibitem{Hanson2007} R. Hanson, L. P. Kouwenhoven, J. R. Petta, S. Tarucha, L. M. K. Vandersypen,  Reviews of Modern Physics, \textbf{79}, 1217 (2007).

\bibitem{PioroLadriere2008} M. Pioro-Ladri\`ere, T. Obata, Y. Tokura, Y.-S. Shin, T. Kubo, K. Yoshida, T. Taniyama, and S. Tarucha, Nature Phys. \textbf{4}, 776 (2008).

\bibitem{Studenikin2012}  S. A. Studenikin, G. C. Aers, G. Granger, L. Gaudreau, A. Kam, P. Zawadzki, Z. R. Wasilewski, and A. S. Sachrajda.  Phys. Rev. Lett. \textbf{108}, 226802 (2012).

\bibitem{StudenikinAPL2012} S. A. Studenikin, J. Thorgrimson, G. C. Aers, A. Kam, P. Zawadzki, Z. R. Wasilewski, A. Bogan, and A. S. Sachrajda, Appl. Phys. Lett. 101, 233101 (2012).

\bibitem{Laird2010} E. A. Laird, J. M. Taylor, D. P. DiVincenzo, C. M. Marcus, M. P. Hanson, and A. C. Gossard, Phys. Rev. B \textbf{82}, 075403 (2010).

\bibitem{Gaudreau2011} L. Gaudreau, G. Granger, A. Kam, G. C. Aers, S. A. Studenikin, P. Zawadzki, M. Pioro-Ladri\`ere, Z. R. Wasilewski, A. S. Sachrajda, Nat. Phys. \textbf{8}, 54 (2012).

\bibitem{Aers2012} G. C. Aers, S. A. Studenikin, G. Granger, A. Kam, P. Zawadzki, Z. R. Wasilewski, and A. S. Sachrajda.  Phys. Rev. B \textbf{86}, 045316 (2012).

\bibitem{Studenikin2013}  S. Studenikin, G. Aers, G. Granger, L. Gaudreau, A. Kam, P. Zawadzki, Z. R. Wasilewski, A. Sachrajda, Phys. Status Solidi C \textbf{10}, 752 (2013).

\bibitem{Medford2013} J. Medford, J. Beil, J. M. Taylor, E. I. Rashba, H. Lu, A. C. Gossard, and C. M. Marcus, Phys. Rev. Lett. \textbf{111}, 050501 (2013).

\bibitem{vanWeperen2011} I. van Weperen, B. D. Armstrong, E. A. Laird, J. Medford, C. M. Marcus, M. P. Hanson, A. C. Gossard, Phys. Rev. Lett. \textbf{107}, 030506 (2011).

\bibitem{Petta2010} J. R. Petta, H. Lu, and A. C. Gossard, Science \textbf{327}, 669 (2010).

\bibitem{Nalbach2013} P. Nalbach, J. Kn\"orzer, and S. Ludwig, Phys. Rev. B \textbf{87} 165425 (2013).




\bibitem{Landau1932} L. Landau, Phys. Z. Sowjetunion \textbf{2}, 46 (1932).

\bibitem{Zener1932} C. Zener, Proc. R. Soc. London, Ser. A \textbf{137}, 696 (1932).

\bibitem{Stuckelberg1932}  E. C. G. St\"uckelberg, Helv. Phys. Acta \textbf{5}, 369 (1932).


\bibitem{Shevchenko2010} S. N. Shevchenko, S. Ashhab, Franco Nori, Physics Reports \textbf{492}, 1 (2010).

\bibitem{Ribeiro2010} H. Ribeiro, J. R. Petta, and G. Burkard, Phys. Rev. B \textbf{82}, 115445 (2010).

\bibitem{Burkard2010} G. Burkard, Science \textbf{327}, 650 (2010).

\bibitem{Sarkka2011} J. S\"arkk\"a and A. Harju, New J. of Phys. \textbf{13}, 043010 (2011).





\bibitem{Gaudreau2009} L. Gaudreau, A. Kam, G. Granger, S.A. Studenikin, P. Zawadzki, and A.S. Sachrajda, Appl. Phys. Lett. \textbf{95}, 193101 (2009).

\bibitem{Field1993} M. Field, C. G. Smith, M. Pepper, D. A. Ritchie, J. E. F. Frost, G. A. C. Jones, and D. G. Hasko, Phys. Rev. Lett. \textbf{70}, 1311 (1993).

\bibitem{Ono2002} K. Ono, D. G. Austing, Y. Tokura, S. Tarucha, Science \textbf{297}, 1313 (2002).

\bibitem{Footnote} Initialization procedures did not achieve a significant improvement in the experimental visibility and we do not show any experimental results in the IS case in this paper.

\bibitem{Taylor2007} J. M. Taylor, \textit{et al.} Phys. Rev. B \textbf{76}, 035315 (2007).

\bibitem{Barthel2012} C. Barthel, J. Medford, H. Bluhm, A. Yacoby, C. M. Marcus, M. P. Hanson, and A. C. Gossard, Phys. Rev. B \textbf{85}, 035306 (2012).



\end{thebibliography}
\end{document}